\documentclass{elsearticle}

\def \eq {\begin{equation}}
\def \qe {\end{equation}}
\def \eqa {\begin{eqnarray}}
\def \aqe {\end{eqnarray}}

\DeclareMathSizes{12}{12}{10}{6}  
\newcommand{\comment}[1]{}
\usepackage[]{graphicx}
\usepackage{caption}
\usepackage{amsmath}
\usepackage{amssymb,amsfonts,textcomp}
\usepackage{caption}
\usepackage{subcaption}
\usepackage{amssymb}
\usepackage{float}
\DeclareCaptionFormat{subfig}{\figurename~(#1)#2#3}
\DeclareCaptionSubType*{figure}
\begin{document}
\begin{frontmatter}

\title{Cross-modal codification of images with auditory stimuli: a language for the visually impaired}
\author{Takahisa Kishino}
\author{Sun Zhe}
\author{R uggero Micheletto}
\ead{ruggero@yokohama-cu.ac.jp}
\address{Yokohama City University, Graduate School of Nanobioscience, Department of Nanosystem Science, 22-2 Seto Kanazawa-ku, 236-0027 Yokohama, Japan}
\author{Roberto Marchisio}, 
\address{Hicare Research S.r.l, Environment Park, Turin, Italy}
\begin{abstract}
In this study we describe a methodology to realize visual images cognition in the broader sense, by a cross-modal stimulation through the auditory channel.
An original algorithm of conversion from bi-dimensional images to sounds has been established and tested on several subjects.
Our results show that subjects where able to discriminate with a precision of 95\% different sounds corresponding to different test geometric shapes. Moreover, after brief learning sessions on simple images, subjects where able to recognize among a group of 16 complex and never-trained images a single target by hearing its acoustical counterpart. Rate of recognition was found to depend on image characteristics, in 90\% of the cases, subjects did better than choosing at random. This study contribute to the understanding of
cross-modal perception and help for the realization of systems that use acoustical signals to help visually impaired persons to recognize objects and improve navigation.
\end{abstract}
\end{frontmatter}
\section{Introduction}
In the latest decades we have seen extreme improvements in technologies related to computers and computer hardware. Processing speed, data transfer speed, memory etc. have improved several order of magnitude. This has led to progress in the field of human interfaces, touch screens, 3D technologies just to name few active fields nowadays\cite{Yoshihiro1996}. 
Despite these advancements, no revolutionary technology have been introduced for people with visual disabilities that we can still see accompanied by dogs, friends, or walking alone with the aid of a stick\cite{kobayashi1997}\cite{Nakamura1997}\cite{Kaluwahandi2001}. Even if several devices have been developed in the effort to help visually impaired people. Notably optical character recognition devices, speech synthesizer or Braille electronic boards\cite{suzuki2003}\cite{tanaka2008}, all of these are aids that give the visually impaired an information generated externally by complex algorithms. This study instead wants to focus on the task of helping visually impaired people to generate their own visual percepts in a natural and general way. 
We propose a cross-modal approach that code elementary visual information to acoustical patterns recognizable by a trained subject that use them to generate internally complex visual percepts. 

The use of sounds to represent visual concept is a delicate problem; auditory-visual stimuli are complex to treat because the cross-modal interaction produce uncertain percepts. It has been reported by Lewald and Guski \cite{lew2004} that the point of subjective simultaneity changes with distance because of temporal disparity, moreover visual field size is influenced by the presence of auditory signals {\cite{buck2010}} that reduce the focus of attention. Also, there are links between auditory selective attention and visual-spatial representation \cite{bomb2010}, and it is well known that in cross-modal stimulation the judgement about a stimulus correlates with judgement on the other one even if physically they are uncorrelated\cite{ziem2013}.
This leads to think that vision is not the dominant sensory modality, self-contained and independent from other senses, many results show that visual perception is strongly altered by other stimuli, as sound or touch\cite{sham2010,tura2005}. Timing between auditory and visual signals is also extremely important, two signals become strongly perceptually bound when they appear simultaneous, even if the source of those signals cannot have occurred together\cite{arno2005}. All of these experimental clues put forward a new hypothetical enlarged definition of {\it objecthood} by which the perception of an object is related to a cross-modal conception that focuses on the similarities between modalities instead of differences as traditionally thought\cite{kubo2001}.

In this framework we devised a particular type of cross-modal approach in which the visual stimuli is actually absent, whereas the visual perception is induced by an auditory signal. Training is used to firstly induce the subject to consolidate a link between two auditory and visual representation of objects through a training session, then we remove one of the modes (visual) and test the ability of the subject to maintain the link and recognize the objects. \par Moreover, we demonstrate that the system works independently of the shape or the geometrical pattern used and we demonstrate the subject's ability to recognize new, never trained shapes. 

\section{System model}
In many cross-modal image recognition software images are scanned for features that are recognized by algorithms. After the processing the original picture is represented by a higher level of abstract patterns that are extracted and then converted in sounds accordingly to semantic rules coded in the output signal\cite{Yoshihiro1996,Kaluwahandi2001,snai1998,molt1998}. In this way the amount of information to be delivered to the visually impaired person is reduced, with various advantages.
Nevertheless, this approach has the disadvantage to limit our analysis to well defined patterns or features, those must be indicated to the software in advance and elaborated accordingly\cite{kish2013}. If a feature is not exactly predictable, variable or affected by noise, the algorithm will fail or miss to transmit an important clue to the visually impaired subject. 

\par
Also, Braille, tongue interfaces and other\cite{Watanabe20002}, try to approach the same problem by converting images detected by a CCD camera to vibrotactile sensations that are understood as geometrical perceptions after a proper training.

These strategies had successful results in specific cases, nevertheless images suffer resolution restriction because tactile senses have very weak spatial discrimination. Also, tactile signals are difficult to superimpose, so multiple images cannot be delivered simultaneously. Transducers to reproduce suitable vibrational signals are not-standard devices, the development of practicable tools is expensive and relevant research takes time and it is difficult.

\begin{figure}[h]
\centering
\includegraphics[width=77mm]{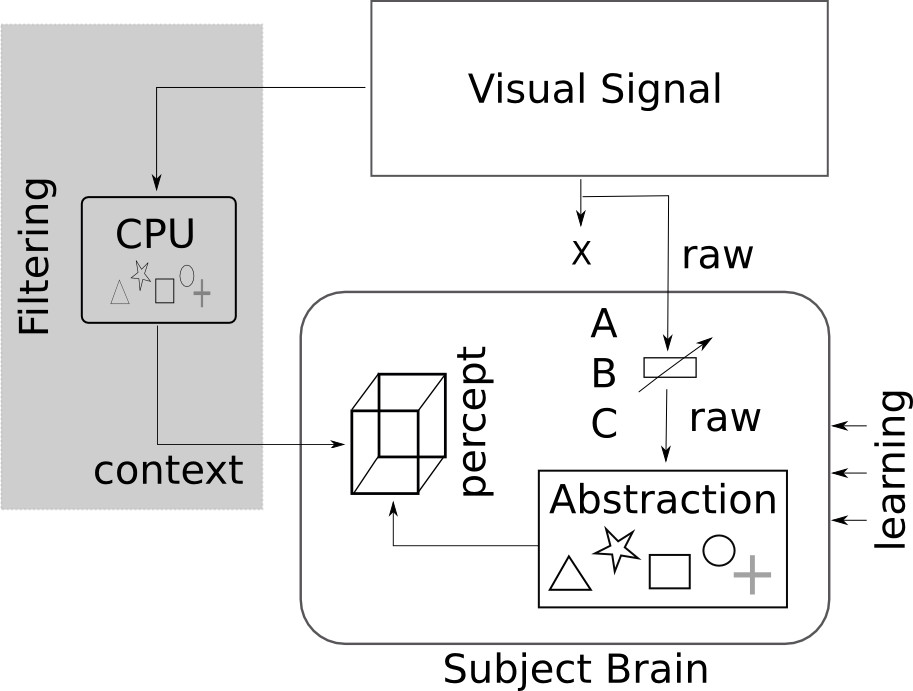}
\caption{A sketch of signal conversion for a cross-modal algorithm for the visually impaired. Two methodologies are shown: greyed on the left is the common procedure where signal is heavily treated externally through a CPU that perform a filtering stage. In there a complex algorithm scans the signal for known features (typically pattern recognition, symbolized by the 5 shapes in the box) that are recognized and sent to the subject as a contextual signal. Notice the {\it subject brain} box and the perception formation symbolized as a cube. The other pathway is our methodology that instead uses raw visual signal converted by a cross-modal algorithm. No external filtering or pattern recognition stage is used, the subject brain perform all abstract processes reinforced by a learning phase. The flow to the "x" mark represents the visually impaired person interrupted visual channel. "A", "B" and "C" are the three cross-modal modalities used in this study and the five shapes group symbolizes the abstraction process that leads to pattern recognition and realizes a perception.}
\label{fig:mp}
\end{figure}

Our visual-acoustic approach does not require external high level abstract models, instead elementary image treatment results are delivered directly to the subject brain by cross-modal conversion. The abstraction and recognition process is done entirely by the subject brain. This strategy has the disadvantage of the need of a learning phase, but has the tremendous convenience of the abstraction ability of the brain, resulting in inherently general, robust and flexible outcome. See figure \ref{fig:mp} for a scheme of the processing procedure.

\subsection*{The algorithm}
Our cross-modal methodology to transfer visual information to the brain, uses sounds that encode the raw data of a picture. Retinal ganglion cells studies have demonstrated that algorithms on the retina pre-filter the physical image prior to be sent to the brain for higher elaboration\cite{russ2010}. In other words, the retina acts as a first stage edge detection processor. Other researches also point to the fact that brains focus on spatial variation of intensity more than absolute local strength\cite{pasu1999,henr1994}. For these reasons, we considered closely to establish an {\it edge detection} elaboration strategy to generate the sounds.
The algorithms we implemented retrieve an image from a CCD camera, then converted to a 8 bit, gray level 200x200 pixels figure. This is filtered for edge recognition, each pixel assigned to be an edge is set to a luminosity of 255 and all the other to zero. 
The image is then scanned from left to right along the horizontal axis in order to generate the cross-modal audio signal. Each of the 200 horizontal columns, is analyzed for its intensity values. As said above, possible values are only 255 ("on") and zero ("off"). If a pixel in the column is "on" a tone is generated (the quality of the tone is determined by a conversion {\it mode} described later). For each column, there are 200 vertical position, so there are 200 possible tones to generate. 
For each image the scan is performed in a total duration of one second. This value allows for each horizontal column an acoustical perception time of $1/200=5$msec, enough to be completely recognized by the subjects.   

\begin{figure}[h]
\centering
\includegraphics[width=77mm]{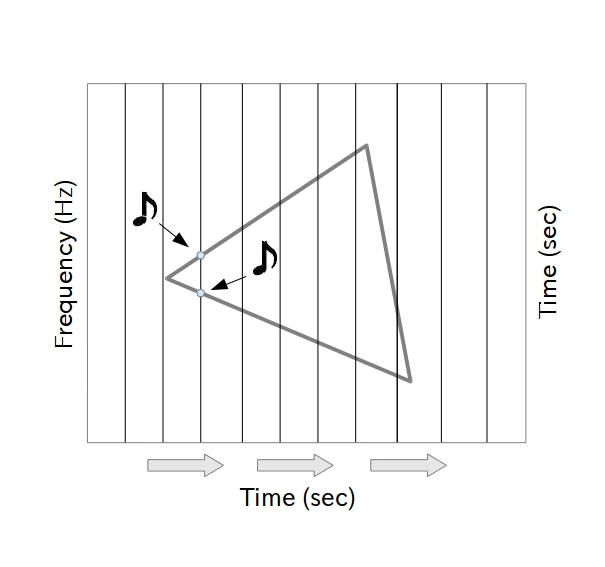}
\caption{A schematic representation of the image scanning algorithm. The triangular shape in the example is scanned from left to right, while a sound is generated accordingly to a conversion {\it mode} that uses the vertical pixel position as parameter.}
\label{fig:TA.2}
\end{figure}

We used initially three test methods to convert the images in sound. All of them were realized using the generic framework where the image is scanned from left to right, each horizontal pixel $x$ addresses a column of $y$ pixels. The luminosity value of this column of pixels is used to generate a specific sound at point $x$.
See figure \ref{fig:TA.2} for a schematic model of the algorithm.

We devised three method of conversion. The first one, called hereafter \underline{method "A"}, consisted in the straight conversion of luminosity to sound intensity. In this method the intensity of each pixel in the column $I_x(y)$ ($x\in\{0..\text{width} \}$ and $y\in\{0..\text{height \}}$, $\text{width=height=}200$ pixels) is directly converted in sound intensity. In other words, the sound wave has the same shape as the luminosity curve $S_x(t)=I_x(y)$. With $S_x(t)$ we represent a curve that for each column $I_x$ have a duration of $t$, and its amplitude is proportional to the luminosity of the image pixels in the column $I_x(y)$. If in this column there are more than two luminous points, say point $y_1$ and point $y_2$, these values are heard as two clicks at times $t_1$ and $t_2$, each of duration t divided by the number of pixels in the column $t/200$. These two clicks will be separated in time by an amount proportional to the number pixels separating each other in the column, $t_2-t_1\propto dN_{pix}$ where $dN_{pix}$ is the vertical distance in pixels of the two points in the column. With the same logic, each dark pixel in the column has a duration of $t/200$ and no sound intensity.  
Scanning the image from left to right gives a total sound wave that is the concatenation of all $S_x(t)$ along the $x$ axis. Since there are in our case 200 pixels, the total duration will be $\tau$=200$*t$ (t is set to be 5msec in order to have a duration of a second), we write $$\Sigma=\bar{\sum}_{x=1}^{x=\text{width}}S_x(t)$$ where the sum $\bar{\Sigma}$ is not intended as an arithmetic total, but as the operation of appending the $S_x$ intensity values to the previous one. This is a straightforward and natural way of converting intensity values of an image to sound amplitude, it is simple and preserve all information. 

The second \underline{method "B"}, consists of calculating the inverse Fourier transformation of each column $I_x(y)$. The inverse transformation converts each column in a wave that is appended to the waves generated by similar conversions in the adjacent columns. If there is a single luminous pixel in the column $I_x$ the inverse Fourier will yield a pure sine wave of frequency proportional to the position of the pixel. If there are more than one, the superimposition of sine waves of different frequencies is generated. The calculation is normalized in order to last $t$=5msec seconds, again for each image we have then a chain of 200 of these for a sound of 1 second duration.

The third \underline{method "C"} uses an indirect generation of frequency, each $I_x(y)$ pixel of the column $I_x$ is tested for luminosity, if a pixel it is found to be "on", a sound is generated with frequency dependent on the hight $y$ accordingly to 

\begin{eqnarray}
I_x(y,t)=&sin(2\pi f(y)t) \\ \nonumber
S_x(t)=&\Sigma_{y=0}^{y=\text{height}}I_x(y,t)
\end{eqnarray}

again $I_x(y,t)$ is calibrated to have a duration of 5msec for each column, $y$ the the pixel height in the image, $f(y)$ is a linear function ranging from 200$H_z$ to 20$KH_z$. Similarly to method "B", if two or more pixels are lit in column $I_x$, the superimposition of two or more sine waves is generated as $S_x(t)$. 

\section{Experiment 1: Sound discrimination ability}
The first goal of these tests is to determine the ability of the various algorithms mentioned above (mode "A", "B" or "C") to produce suitable sounds, and to verify how much a human subject is able to distinguish and interpret cognitively the properties of those.
In particular, all the trans-coding procedures begin with an edge filtering, so the images used were simple geometrical images with clear edges. Also, once established if the subject is able to recognize a determined shape by hearing its trans-coded sound, we wanted to test if he/her was aware also of inclination, and other rotational information contained in the sound.   

\begin{figure}[h]
\centering
\includegraphics[width=77mm]{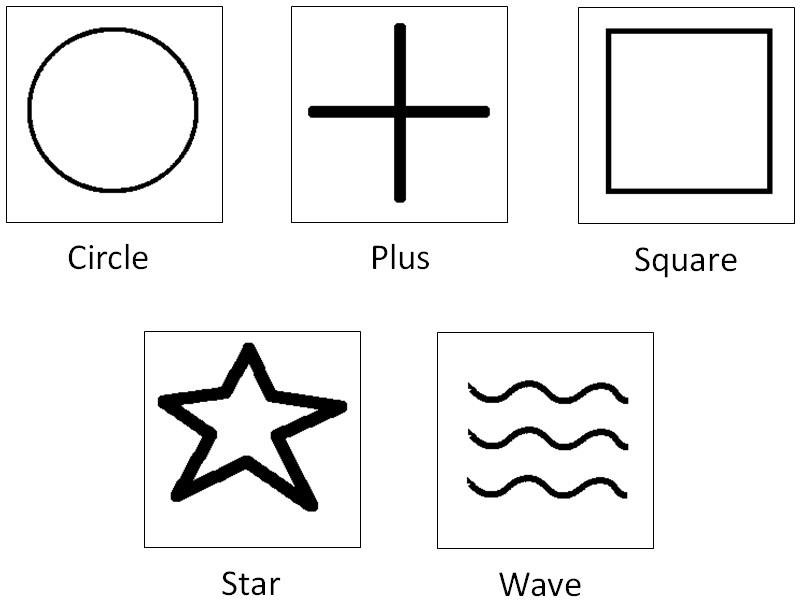}
\caption{The five simple images used for conversion. Each of those pictures are converted to sound accordingly to one of three algorithms of conversion called here mode "A", mode "B" and mode "C".}
\label{fig:TTR.01}
\end{figure}

Initially, the sounds were played to verify if sounds trans-coded from various images were perceived differently by subjects. In other words, in these first experiments, subjects were tested for their ability to discern if two sounds corresponding to two distinct shapes were different or confused. The test did not want to check the capability to recognize the sound and associate it with geometrical entities.
  
\subsection{Method:} 
The subjects used in these tests were 5 adult males of age ranging from 20 to 24 years. None of them had problem of vision nor hearing was impaired. Three of them were playing instruments as a hobby. 

\begin{figure}[h]
\centering
\includegraphics[width=120mm]{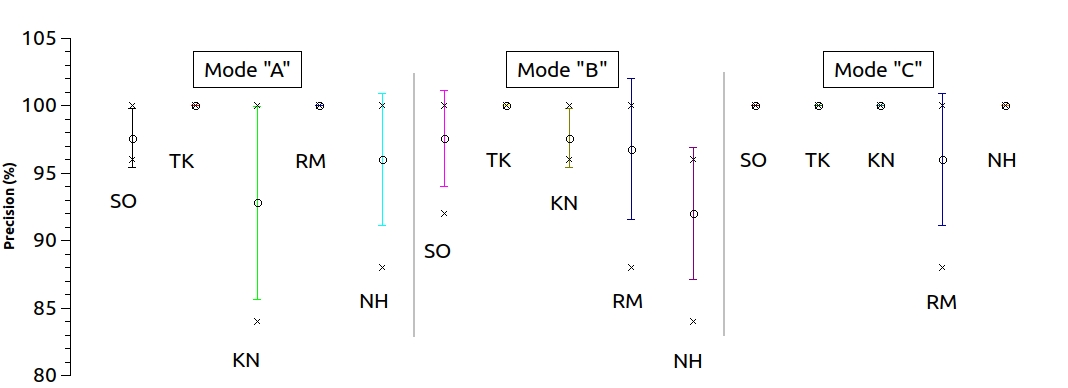}
\caption{The results of discrimination tests. Five subjecs SO, TK, KN, RM and NH are presented with pair of sounds trans-coded from images of simple geometrical shapes (see figure \ref{fig:TTR.01}). The subjects have to tell if the two sounds are coincident, different or if they are not sure. Three trans-coding methods are used, called here mode "A", mode "B" or mode "C". In the case of mode "C" the results show nearly 100\% precision. Each subjects does 5 sessions for each trans-code modality ("A", "B" or "C"), there are 5 tests for each sessions, so each subjects does a total of 25 tests per modality. The circles in the graph represent the median, the crosses the minimum and maximum values and the whiskers the standard deviation range. If a subject perform multiple sessions at 100\% precision, all data appear superimposed.}
\label{fig:TTR.1}
\end{figure}

In figure \ref{fig:TTR.01} are shown the 5 schematic images used. The images were transcoded in real time during the experiment using the three modalities "A", "B" and "C" described above. We developed a software, in C language, that presents to the subjects two sounds corresponding to two images of the five in figure, at random and with possible identical sounds in the pair. The subjects do not know if the sounds correspond to different images or the same images, and after hearing the two sounds in sequence he have to choose between three options: to indicate if the sounds were the same, different or if they were not sure.

The subjects write on a keyboard on a computer text based console. For each of the three transcoding modes "A", "B" and "C", there were 5 sessions with 25 trials (one trial is a set of two sounds played) per sessions. The subjects were not aware of the algorithm functioning, nor they knew the number or the shape of the original 5 images that generates the sounds.  
As shown in the raw data plotted in figure \ref{fig:TTR.1} and in the averaging of table \ref{tab:dis}, the trans-coding algorithm "C", based on frequency translation of vertical pixels, results to be the one more easy to recognize, and the one that gives less probability of error.

\begin{table}[ht!]
\begin{center}
\begin{tabular}{| c | c | c | c |}
\hline
$~$ & mode "A" & mode "B" & mode "C" \\
\hline
Average discrimination precision & 96.80\% & 96.80\% & 99.2\% \\
Standard deviation & 4.8\% & 4.23\% & 2.52\% \\
\hline
\end{tabular}
\end{center}
\caption{Average accuracy of discrimination for the three algorithms used.}
\label{tab:dis}
\end{table}

\section{Experiment 2: Association between sounds and images} 
Instead to investigate the ability of the subjects to recognize the sound differences, in this experiment we focused on the actual association between sounds and specific geometrical images.
The subjects are presented an image and hear the sounds corresponding to the conversion modality used in the session. The subjects try to detect particular acoustic clues and associate them with visual clues on the image. Since all modalities involve edge recognition, these clues can be inclination of lines, corners or curved shapes for example. These visual clues are spontaneously associated with acoustical patterns creating a cross-modal perception learning process.
For every modality the subjects are presented the five images mentioned above, simultaneously in one panel, and one sound corresponding to the conversion of one of these. The subject have to choose one image that he/her thinks is the most probably associated to the sound. Data are collected and statistical data are recorded with the distribution of the associations. The test is repeated similarly for the three modalities. 

\subsection{Method}
The subjects that participated in the experiment were four adults. None of them have any hearing problems, two of them have the hobby of playing an instrument.
Since five different sounds stimuli were used with the three conversion modality, in total the subjects are exposed to 15 different sound stimuli.

When the sound is played the subject does not know at what image it is associated, but he is aware that it is one of the five pictures in figure \ref{fig:TTR.01}. There is no learning session, so the subjects do not memorize sound and images, but instead try to associate sounds clues with visual details. The subjects are not aware of the functioning of the conversion algorithms, nor of their differences. Every sound is extracted randomly from the set and the algorithm presenting the sounds will repeat the same stimulus five times. There are three sessions corresponding to the three conversion modalities described in the previous chapters for a total of 75 questions per session.

\subsection{Results and discussion}

Before making comments about the tests we have to consider some differences between the five optical stimuli. The cross and the square in figure \ref{fig:TTR.01} have a simple right angle structure, so the edge detection phase will be limited to a small area of the image. The other three images, the star, the wave and the circle instead are composed by curves or inclined lines. This will results in higher importance of the edge detections on a wider area of the image, compared to the previous two stimuli.

\begin{figure}[h]
\centering
\includegraphics[width=77mm]{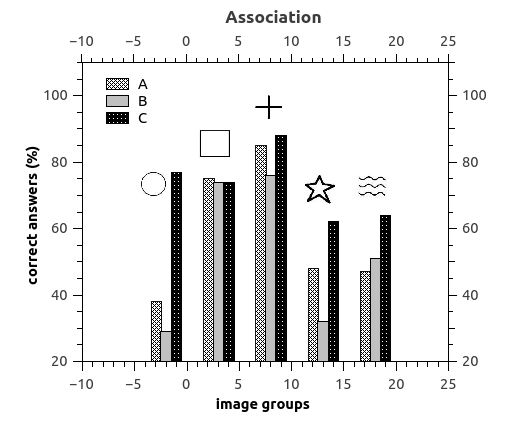}
\caption{Association tests were done using the five simple shapes shown. Each group of bars represents the correct association between the sound stimulus and the shape from which it was generated. The three bar in each group indicate the different sound conversion methods "A", "B" or "C". Overall, the method "C" seems to report better results for any image.}
\label{fig:M.1}
\end{figure}

The results of the test for all the three method of conversion are reported in figure \ref{fig:M.1}.

It is noticeable that with method "A" the circle and the other two most curved images, the star and the wave, have lower values. This is due to the nature of this specific conversion method, that as stated above, directly links the pixel intensity of the image to the final wave. Shapes with steep optical variations like the square or the cross result in better association rates with this modality.

Using the conversion mode "B" a Fourier transformation is involved in the cross-modal process. The results of the tests show a basic agreement with the direct conversion in modality "A" above. This result is related to the fact that Fourier transformation is done to every single column $I_x$ of the image, not on the whole image. The experimental study of this method have been driven by its natural affinity with general Brain activities, where high level processing always implicate Fourier transformations. Nevertheless, the actual exercise of the experiments shows no particular advantage against a simpler conversion algorithm like method "A".

The tests show clearly better association results for the more complex mode "C", in which synthetic frequencies are generated in function of the pixel position $I_x(y)$, as described above.Not only the images that present higher curvature like the circle or the wave perform with better association results, but also the images characterized by straight horizontal lines like the cross or the square give rise to better results. Overall this test show that the modality "C" give more reliable association power.

\section{Learning ability analysis}

In the previous tests described above, we understood that the sounds corresponding to images with different characteristics are indeed different and they are perceived with mixed precisions. Conversion modes influenced the recognition probability to a certain amount, however, our results demonstrates also that particular visual characteristics of an image, sharp corners, inclination, curves etc., are tied to the acoustical representation of them, independently from the conversion modality used. This is an important consideration to take in account in the phase of learning.

If, after a suitable learning session on a fixed number of standard images, a subject becomes able to associate particular sound characteristics, to specific image details, then they will be able generalize the recognition process to associate new sounds to images that they do not know and that they never have been trained on. 
The goal of the learning process described hereafter is to verify and characterize quantitatively the implications of this hypothesis\cite{Holroyd2002,Shiffrin1977,John2000}.

\subsection{Experiment 3: orientation of linear details}
\begin{figure}[h]
\centering
\includegraphics[width=77mm]{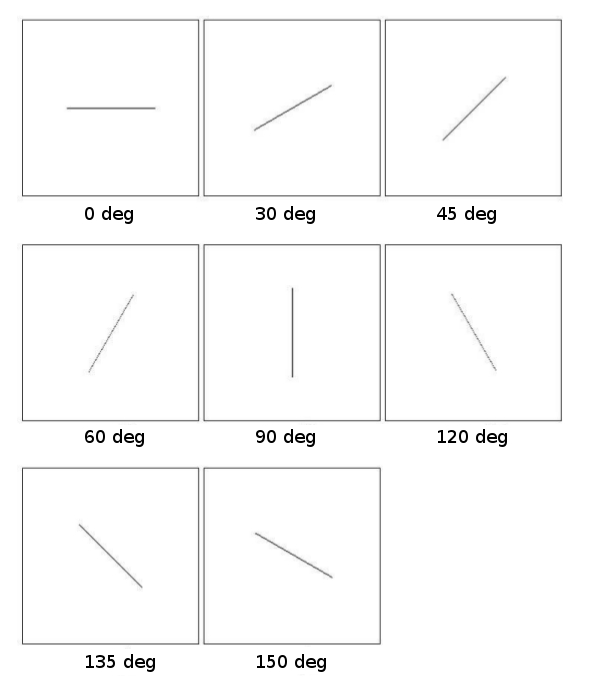}
\caption{The image samples used in the tests with elemental visual stimuli. Every sample is a linear segment with the same length and different inclination.}
\label{fig:LOLD.1}
\end{figure}

In this section we want to investigate the ability to recognize the orientation of edges or other linear elements of an image. 
This study is of fundamental importance because any complex geometrical object is a composition of simpler shapes that can be decomposed in segments. If subjects are able to recognize the orientation of a straight line by hearing a sound, then we expect they will be able to recognize complex images.
In this test, 8 segment of fixed length with different inclination angles were used (as in figure \ref{fig:LOLD.1}).

 The images were presented to subjects in random order, they were wearing headphones and could hear the sound corresponding to each of them accordingly to the three conversion methods "A", "B" or "C". The subject select a conversion method, and in random order one-by-one the eight segments sounds are played. In this learning phase the subjects knows what segment inclinations the sound corresponds to. Subjects are allowed to hear this sound as many times as necessary. Once the subject feels satisfied, he can proceed with the next random image until all 8 of them are done. The subjects were not allowed to hear again a sound corresponding to a previous sample, they were only able to repeatedly hear the same sound for a particular image, but once they are done with that, they had to proceed to the next one without the possibility to rehearsal on previous images. Usually this process takes about 5-10 minutes for each conversion method. So this learning phase can last between 15 minutes to half an hour.

Once the learning procedure is finished, the subjects start with the recognition tests. Without any image being shown, the subject hears 6 sounds corresponding to one of the images in figure \ref{fig:LOLD.1}, extracted at random. For each of these sounds, the subjects are asked to estimate the inclination of the segment and to draw the image they think it is corresponding on a piece of paper. Also in this case, the subjects were allowed to re-hear the sound for that particular image. Once the answer is drawn, the subjects were requested to proceed to the next sound. When all the six tests have been done with one conversion modality, the subjects proceed to tests with next one, with other six images extracted at random from fig. \ref{fig:LOLD.1}. These tests last about another half an hour in total for the three modalities.

Each subject does this learning-test session once or twice a week for a total of ten sessions in the span of about two months. Five subjects were chosen among students or colleagues, none had known problems of eyesight or hearing. Subject were given the equivalent of 30 US dollars, as a small compensation for participating to the study.

\subsection{Results and discussion}
\begin{figure}[!ht]
\begin{minipage}[b]{0.5\linewidth}

  \centering
{\includegraphics[width=75mm]{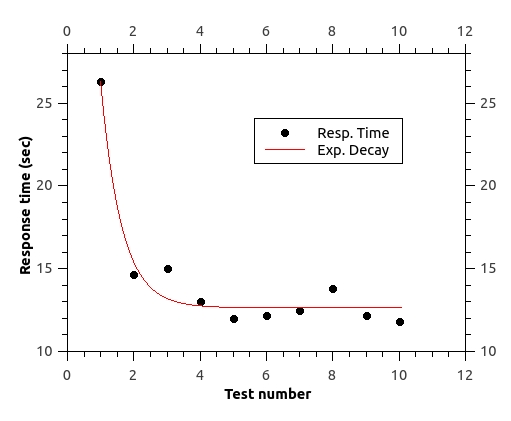}}
\caption*{(a)}
\end{minipage}
\begin{minipage}[b]{0.5\linewidth}
{\includegraphics[width=75mm]{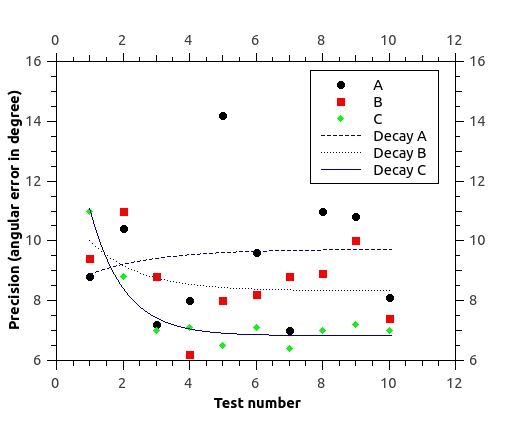}}
\caption*{(b)}
\end{minipage}

\caption{The figure on the left shows how the response time drops exponentially with the test number for all conversion modes averaged. However, on the right is shown the precision for each mode, and mode "C" is the more efficient since the error drops exponentially faster than with the other modalities. Each point in \ref{fig:LOLD.2}a is the average of 6x3 tests for 5 subjects, for a total of 90 tests each point. In figure \ref{fig:LOLD.2}b there are 30 tests for each points.}
\label{fig:LOLD.2}
  \end{figure}

In the left panel of figure \ref{fig:LOLD.2} it is shown the average response time against the number of tests done. The response time is measured taking the time since the sound is heard, until the subject press the button to hear the next sound corresponding to the next segment. 

This is a reliable measure of the average response time since subjects do not take pauses during sessions, but only between measurements sessions. In the case of the last image presented, the response time is taken when the subject press the "end" bottom as instructed. During the tests, a supervisor (K.T.) is present and observe in silence the subjects at work. 

  There is an evident decrease of the response time independently to the cross-modal modality used. This is interesting, and suggests that the subjects become able to take a decision about the segment inclination quickly after few tests for every conversion mode used.
  
  However, in the right panel are the error by which subjects estimate the segment inclination. It is noticeable that quicker response time does not correspond to better precision for all modes. 
  Mode "C" results to be the most precise, the error in angular degree is dropping with test number in an exponential fashion. For this mode, the average of all subjects shows that after about 3-4 tests the error by which subjects estimate the segment inclination reaches a minimum of about 7 degree.
  The inclination angle reported by the subjects were estimated analyzing the drawing they reported for each sound. Since subjects draw the segments not perfectly straight, the inclination was estimated measuring the angle at both ends of the segment and making an average.
In figure \ref{fig:LOLD.2} is also noticeable a slight increase in angular error with test number. This effect is not clear, but we think it may be related to fatigue.

\par

\begin{figure}[h]
\centering
\includegraphics[width=77mm]{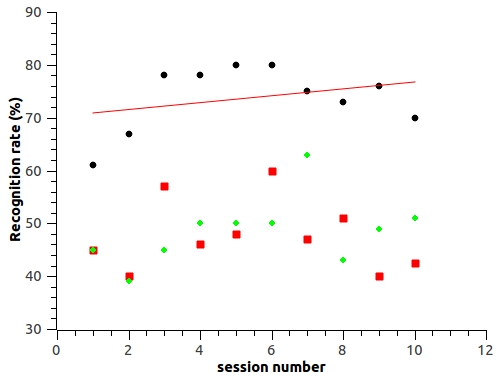}
\caption{The recognition rate in function of session number for each conversion mode. Diamonds (green), squares (red) and dots (black) represent mode "A", "B" and "C", respectively. Mode "C" performs better and shows improvement with sessions. The red line is the best linear fit for this mode, it shows an average 0.7\% recognition rate improvement for each session.}
\label{fig:LOLD.3}
\end{figure}

The drawing made by the subjects were evaluated one by one using a arbitrary but consistent method. Each draw was given 1 point if the inclination sign was recognized and another point if the angle of inclination was within $\pm 5^\circ$ of the correct value. Each session consisted of 6 tests for a total of possible 12 points per session. Each subjects were exposed to 10 sessions, the total result was averaged and normalized to 100$\%$. We call "recognition rate" this average. In figure \ref{fig:LOLD.3} it is shown the dependance of the recognition rate with session number, for all subjects averaged. It is noticeable how conversion modality "C" has the best recognition rate (average about 74$\%$) compared to the other modes ("B" about 48$\%$, "A" about 49$\%$). Mode "A" and "C" show an improvement in recognition rate with test number, whereas modality "B" seems to maintain its low rate independently to that.

\section{Experiment 4: recognition of untrained complex images}

We think that when a subject is able to recognize a shape by its cross-modal conversion sound, they will be able to recognize the shape of an unknown one too, by comparing known sound characteristics with the new one. In this experiment the subjects listen to sounds generated from basic simple shapes and their are trained to recognize them. Then the subjects are exposed to new sounds associated to unknown images. The key to the recognition of never trained images is the ability to recognize differences of the new acoustical signal with previously learned signals relative to known images. The abstract generalization of these differences produces the perception of the new image. For example a sound could be very similar to a known one, but of longer duration. The subject may think of a bigger or larger object compared to the original one. This experiment is focused to understand and possibly demonstrate the fact that untrained images can be recognized and to analyze each conversion mode in this sense.

\subsection{method}

\begin{figure}[h]
\centering
\includegraphics[width=77mm]{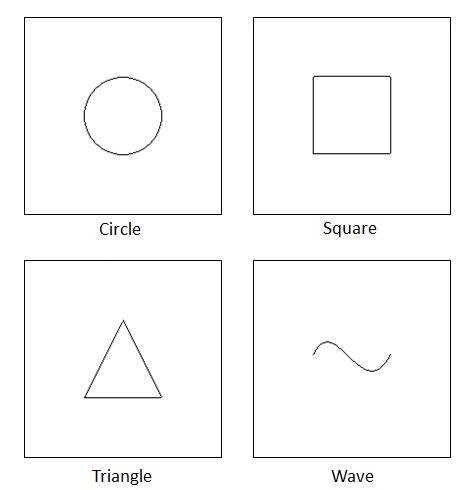}
\caption{The four images used in the learning phase of the experiment. The subjects were trained with cross-modal signals generated by three algorithms from these images.}
\label{fig:RUCI.1}
\end{figure}

The first part of the experiment consists in an initial training where the subjects were exposed to 4 sounds relative to the 4 basic shapes in figure \ref{fig:RUCI.1}. These pictures were given to the subjects in advance, and visible on a print sheet during the experiment.

\begin{figure}[h]
\centering
\includegraphics[width=77mm]{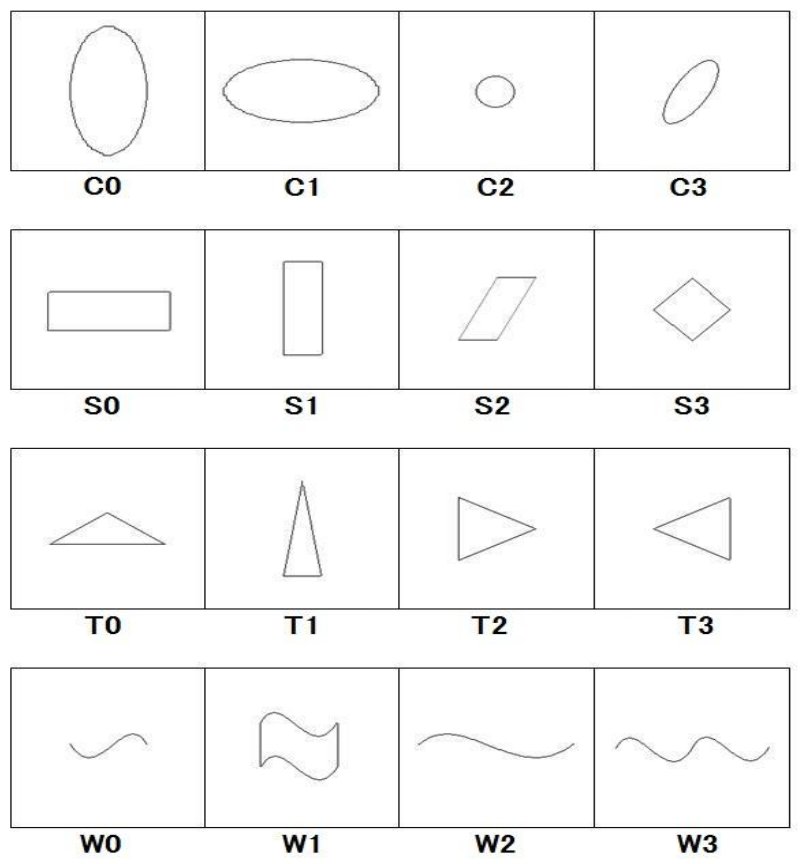}
\caption{Sixteen images used as not-trained cross-modal stimuli. These are modified version of the four trained basic shapes in figure \ref{fig:RUCI.1}.}
\label{fig:RUCI.2}
\end{figure}

 The sounds were presented in random order, with clear indication of what picture the sound is corresponding to. Each sound could be repeated until the subject is satisfied, in the same fashion as in previous tests. No questions were asked at this stage. After this, the subjects were presented to four new sounds relative to a modified version of the basic shapes chosen from a pool of 16 new never trained shapes. They were asked to write on a piece of paper what they thought it was the corresponding shape. Again, they were allowed to hear at will the sound stimulus. In figure \ref{fig:RUCI.2} are shown the 16 possible variations we used. The subjects knew that the new sounds were different from the four that they just heard, however they were not aware of the existence of these 16 variations, nor they were told that the sounds stimuli were chosen from a fixed set. Subjects were simply asked to write down what what they though it was the original picture.

At the end of the four problems, the subjects were shown the answers, with the images relative to the sound stimuli they were presented. This process was repeated three times for the three conversion modes "A", "B" and "C". The subjects at this points were dismissed and experiment closed. Usually a complete experiment lasted about 20-40 minutes.

These experiments were repeated in different days 8 times for each subjects. So each of the five subjects was doing a total of 4x3x8=96 single tests. Sounds were heard through high fidelity headphones, the subjects were operating a computer that was generating the sounds at a press of a key.

\subsection{Results and discussion}

As in previous experiments, we needed a standard method to analyze and evaluate quantitatively the subjects answers. The following evaluation method was chosen: if the subject was able to recognize the primary characteristics of the image (elongated, rotated, etc) one point was assigned, if the image was exact (correct shape, and correct characteristic) two points were given. 

\begin{figure}[h]
\centering
\includegraphics[width=77mm]{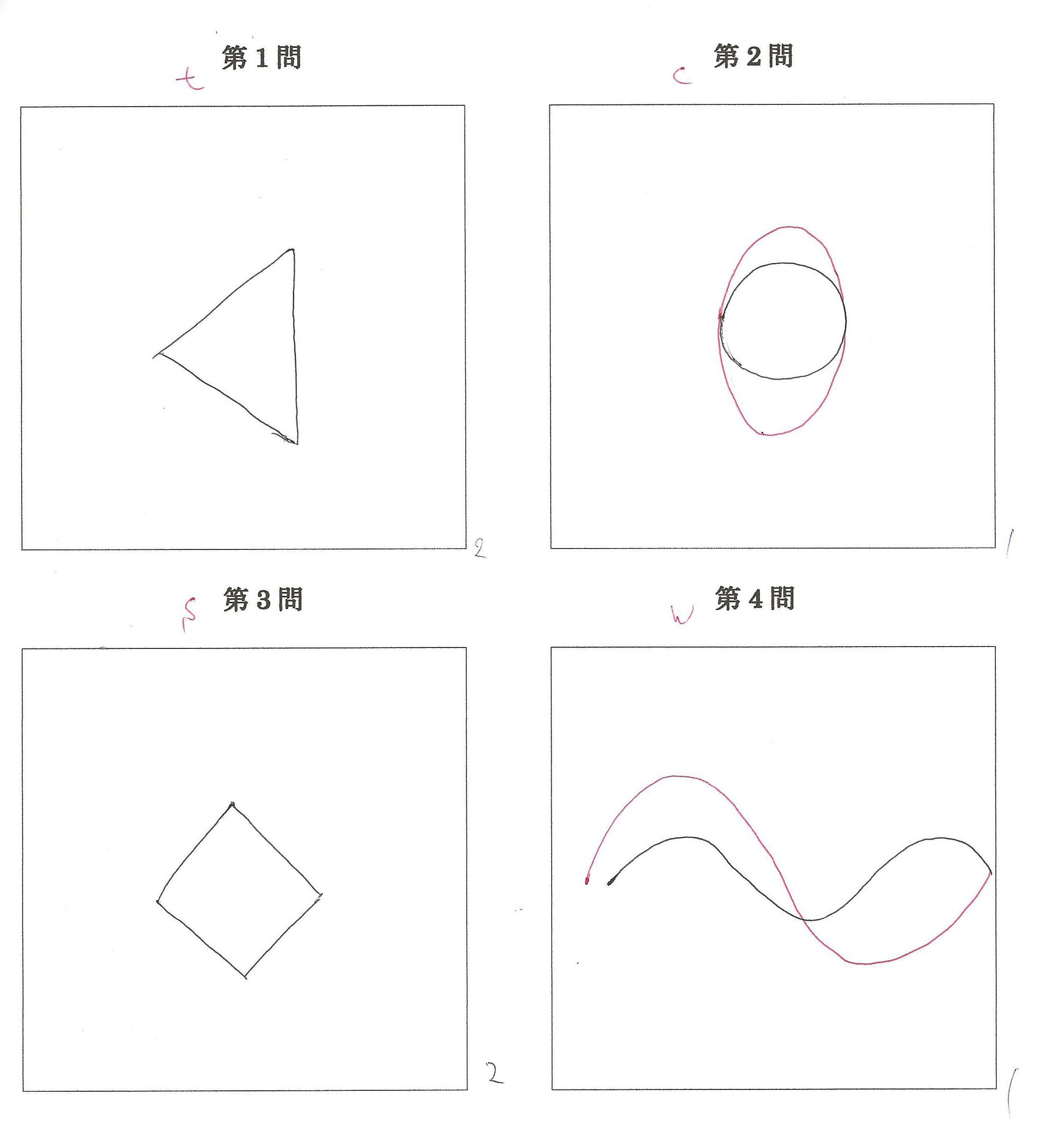}
\caption{An example test report scanned from a real sheet used by a subject. The pictures in blue ink are hand drawn by the subjects and represent what they think is the shape corresponding to the sound heard. The shapes in red are drawn by the referee. Outside the box are reported referee comments in red, and in the right corner in blue the final points assigned to each test (2 points correspond to 100\% in the plot of figure\ref{fig:RUCI.3}.}
\label{fig:RUCI.2.5}
\end{figure}

In figure \ref{fig:RUCI.2.5} is shown one example of the report sheets filled by the subjects. In the square is shown what the subjects wrote, outside it in red are comments of the referee (author T.K.) and in blue in the bottom corner the points assigned using this criteria.

\begin{figure}[h]
\centering
\includegraphics[width=90mm]{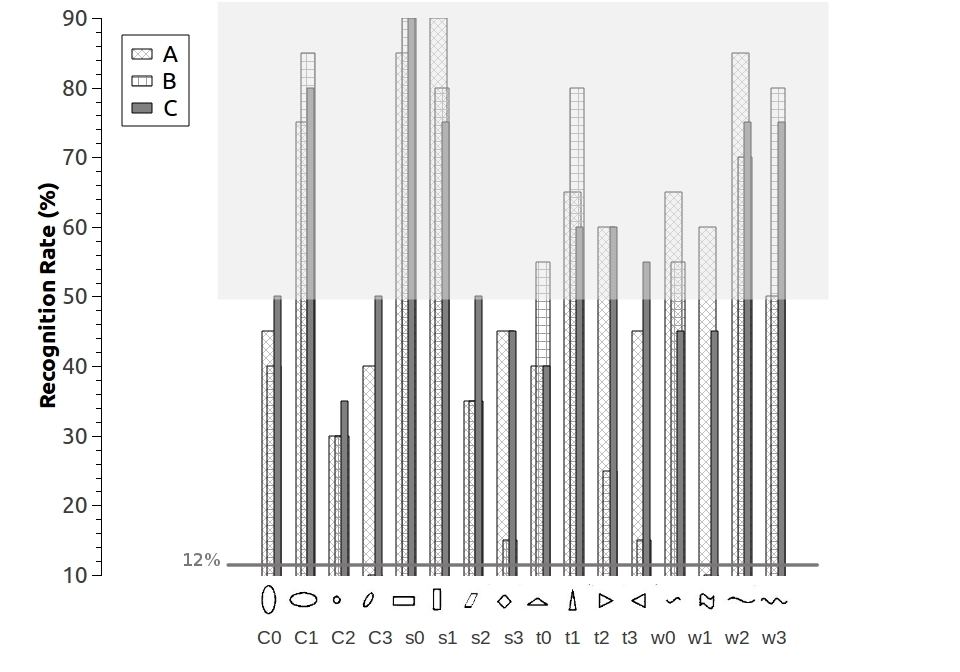}
\caption{The recognition rate of untrained images. Images drawn by the subjects were assigned 0, 1 or 2 points, in this graph 100\% recognition rate means an average of 2 points for each subject. Each subjects gave a total of 96 drawings in eight experiments taken in different days. The grey area emphasizes results better than 50\% recognition rate, whereas the 12\% line at the bottom represents the theoretical result level for randomly selected answers. More than 90\% of the images were recognized with rates higher than this level.}
\label{fig:RUCI.3}
\end{figure}

Overall, taking in account that 2 points correspond to 100\%, the complete results plot is given in figure \ref{fig:RUCI.3}.

Averaging all the 16th images presented to the subjects, for each conversion modality we obtained a recognition rate of 57 \% for mode "A", 48.44\% for mode "B" and 58.13\% for mode "C". We have to notice that these values are far better than choosing the answer at random, that, on this scale, would result in a rate lower than 12\%. Even the images that were more difficult to understand, obtained recognition rates well over the random level.
Figure C1, S0, S1, W2 and W3 performed relatively well in all conversion modes with recognition rates over 70\%.
In the case of images S0 and S1, these are composed by parallel lines. Accordingly to the results of previous experiments as those in chapter 4.2, figure \ref{fig:M.1}, these images are better recognized in all modalities. One of the reason for this good results is that the duration of the sound corresponding to the horizontal line is associated to the length of the object in the image. The four images C1, S0, W2 and W3 are longer horizontally, the subjects could associate the horizontal width of these objects by the duration of the sound in each of the three cross-modal modalities, while the other specific characteristics of the image were deduced easily by comparison of other sound clues with the main four training shapes\cite{Meijer1992}.

\begin{figure}[h]
\centering
\includegraphics[width=77mm]{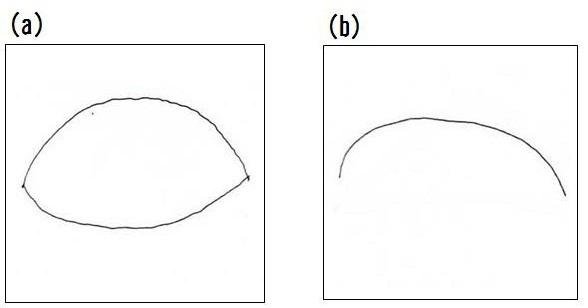}
\caption{An example answer hand drawn by subjects in response to the cross-modal conversion of image T0. This image as a limited vertical extension, so the inclined segments of the triangle were often confused with smoother curves. (a) corresponds to conversion mode "A" and (b) to conversion mode "C".}
\label{fig:RUCI.4}
\end{figure}

The image T0 was often confused to a semicircle or similar image, see actual subjects drawing in fig \ref{fig:RUCI.4}. T0 is a triangle with larger base and limited height, so the smaller inclination of the two inclined segments made them confuse with smooth rounded curves.

\begin{figure}[h]
\centering
\begin{minipage}{.5\textwidth}
  \centering
  \includegraphics[width=77mm]{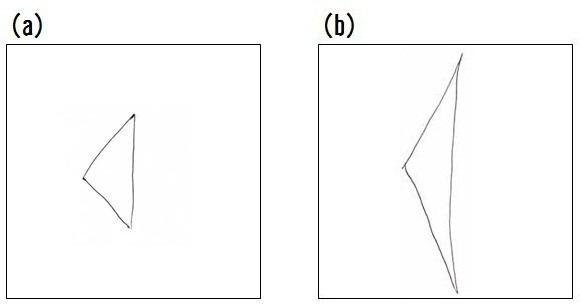}
  \captionof{figure}{A typical mistaken answer relative to image C2 as drawn by the subject. (a) corresponds to conversion mode "A" and (b) to conversion mode "C".}
  \label{fig:RUCI.5}
\end{minipage}%
\quad
\begin{minipage}{.5\textwidth}
  \centering
  \includegraphics[width=77mm]{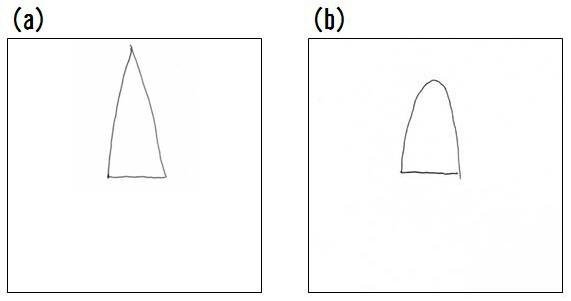}
  \captionof{figure}{Another example of not correct answer relative to image C3. Limited lateral span result in shorter sound duration and this often confuses the subjects, (a) corresponds to conversion mode "A" and (b) to conversion mode "C". See text for discussion.}
  \label{fig:RUCI.6}
\end{minipage}
\end{figure}

On the other hand, images that were horizontally short, like for example C2 and C3, were also often mistaken as shown in figure \ref{fig:RUCI.5} and \ref{fig:RUCI.6}. In all conversion modalities, horizontal span is converted in duration, so these kind of objects were correctly estimated as shorter horizontally, but the detailed vertical structure was often missed. Presumably this was caused by the abrupt change in signal implying less time available for evaluation.

Another image that had relatively low recognition rates is S3, that is small both in height and width so was naturally more difficult for the subjects to analyze.  
\par
\begin{figure}[h]
\centering
\includegraphics[width=75mm]{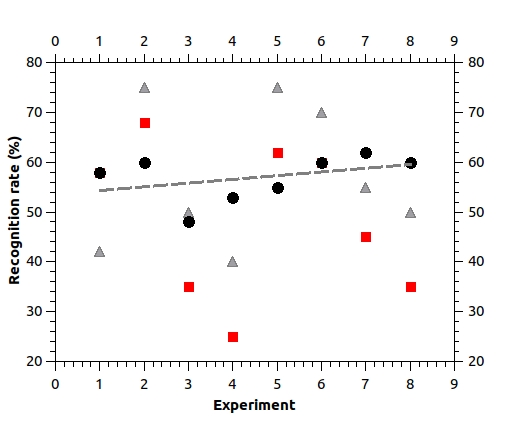}
\caption{The average recognition rate versus the experiment number. The tests for all the images and all the five subjects are averaged for each experiment. The experiments lasted about 30 minutes, a first training session with 4 basic images and then 4 tests of untrained sound stimuli. This was repeated for each conversion mode, so each point in the graph represents the average of 60 tests. The answer were drawn on paper sheets as shown in figure \ref{fig:RUCI.2.5}. Each experiment was performed in 8 different days, within about one month. Mode "A", mode "B" and mode "C" are represented by triangles, squares and circles respectively. Mode "C" results show recognition rate improvements with experiments. The dotted linear interpolation has inclination of about $\alpha_C=0.7$\% per experiment and is calculated over a total of 4x3x8x5=480 tests. Random answers (a complete absence of recognition ability) will result in a level lower than 12\% in this scale.}
\label{fig:RUCI.7}
\end{figure}

In figure \ref{fig:RUCI.7} is shown the average recognition rate averaged on all samples and all subjects in function of experiment number. The linear approximation of data sampled at each test show a quasi monotone positive inclination for method "C", with a coefficient $\alpha_C$=0.7\% per experiment. This shows that the repetition of tests improve learning and this results in better recognition ability. This effect is particularly strong for conversion method "C". Repetition seems to results in better learning and acquisition of better distinction ability.

\subsection{Conclusions}
In this study we investigated the cross-modal conversion of visual images in sounds. We used three different strategies based on a left-to-right raster scan. In the first part of this study, subjects were found to be able to recognize five different images just by hearing their converted sound with a recognition rate better of 95\% for all three modes. \par 
This first experiment demonstrated that the cross-modal approach chosen are reliable for discriminating the acoustical differences involved in the image conversion process of different shapes. 
In a second experiment we studied the subject ability to discriminate inclinations of segments by their acoustical cross-modal conversion. Also in this case we could demonstrate excellent performances, especially for the frequency modulated conversion mode "C" that have shown average angular discrimination error less than 10 degrees.
\par
Finally, we tested the recognition performance on elementary complex images like circles, crosses, waves etc and demonstrated that subjects could be trained easily on these images, and that they recognize even previously untrained images. Results on a range of 16 unknown shapes were much better than choosing the answer at random and this was true for all conversion modes. Again the frequency modulation based mode "C" was the best conversion mode.   

We demonstrated that trained subjects can recognize different images basing themselves only to the cross-modal acoustical information. Also, we have shown that after an appropriate training, subjects are able to infer correctly the shape of an unknown previously untrained images. By comparing sound differences with trained samples, subjects are establishing an abstract acoustical language by which visual percepts are achieved in a generalized manner.

This study is useful for the fundamental understanding of perception and cross-modal processes, but also we believe can be of inspiration for the development new devices for visually impaired people.


\clearpage

\begin{thebibliography}{25}
\expandafter\ifx\csname natexlab\endcsname\relax\def\natexlab#1{#1}\fi
\expandafter\ifx\csname url\endcsname\relax
  \def\url#1{\texttt{#1}}\fi
\expandafter\ifx\csname urlprefix\endcsname\relax\def\urlprefix{URL }\fi

\bibitem{Yoshihiro1996}
Y.~Kawai and F.~Tomita,
\newblock Journal of the Institute of Image Information and Television
  Engineers {\bf 51}, 325 (1996).

\bibitem{kobayashi1997}
M.~Kobayashi and M.~Ohta,
\newblock Society of Biomechanisms (SOBIM) Japan {\bf 21}, 39 (1997).

\bibitem{Nakamura1997}
K.~Nakamura, Y.~Aono, and Y.~Tadokoro,
\newblock Systems and Computers in Japan {\bf 28}, 36 (1997).

\bibitem{Kaluwahandi2001}
K.~Sasadara and T.~Yoshiaki,
\newblock Journal of the Institute of Image Information and Television
  Engineers {\bf 55}, 1499 (2001).

\bibitem{suzuki2003}
S.~Yuji {\em et~al.},
\newblock IEEJ Transactions on Electronics, Information and Systems {\bf 102},
  13 (2003).

\bibitem{tanaka2008}
T.~Makoto and G.~Hideaki,
\newblock Information Processing Society of Japan :Computer Vision and Image
  Media {\bf 115}, 125 (2008).

\bibitem{lew2004}
J.~Lewald and R.~Guski,
\newblock Neuroscience Letters {\bf 357}, 119 (2004).

\bibitem{buck2010}
P.~B. David~Buckley, Charlotte~Codina and O.~Pascalis,
\newblock Vision Research {\bf 50}, 548 (2010).

\bibitem{bomb2010}
M.~D. Bomba and A.~Singhal,
\newblock Brain and Cognition {\bf 74}, 273 (2010).

\bibitem{ziem2013}
A.~N. Michal~Ziembowicz and P.~Winkielman,
\newblock Cognition {\bf 129}, 273 (2013).

\bibitem{sham2010}
L.~Shams and R.~Kim,
\newblock Physics of Life reviews {\bf 7}, 269 (2010).

\bibitem{tura2005}
V.~M. Massimo~Turatto and C.~Umilta,
\newblock Cognition {\bf 96}, B55 (2005).

\bibitem{arno2005}
A.~J. Derek H.~Arnold and S.~Nishida,
\newblock Vision Research {\bf 45}, 1275 (2005).

\bibitem{kubo2001}
M.~Kubovy and D.~V. Valkenburg,
\newblock Cognition {\bf 80}, 97 (2001).

\bibitem{snai1998}
M.~Snaith, D.~Lee, and P.~Probert,
\newblock {IMAGE AND VISION COMPUTING} {\bf {16}}, {225 ({1998}).

\bibitem{molt1998}
N.~Molton, S.~Se, J.~Brady, D.~Lee, and P.~Probert,
\newblock {IMAGE AND VISION COMPUTING} {\bf {16}}, {251 ({1998}).

\bibitem{kish2013}
{Kishino, T and Zhe, S and Micheletto, R},
\newblock A fast and precise hog-adaboost based visual support system capable
  to recognize pedestrian and estimate their distance,
\newblock in {\em New Trends in Image Analysis and Processing – ICIAP 2013},
  edited by A.~Petrosino, L.~Maddalena, and P.~Pala, , 4 Vol.~16, pp. 251--263,
  PO BOX 211, 1000 AE Amsterdam, Netherlands, 2013, Elsevier Science BV.

\bibitem{Watanabe20002}
W.~Tetsuya and K.~Makoto,
\newblock IEICE Technical Report Japan {\bf 100}, 27 (2000).

\bibitem{russ2010}
T.~L. Russell and F.~S. Werblin,
\newblock {Journal of Neurophysiology} {\bf {103}}, {2757 ({2010}).

\bibitem{pasu1999}
A.~Pasupathy and C.~Connor,
\newblock {JOURNAL OF NEUROPHYSIOLOGY} {\bf {82}}, {2490 ({1999}).

\bibitem{henr1994}
G.~Henry, A.~Michalski, B.~Winborne, and R.~McCart,
\newblock {Progress in Neurobiology} {\bf {43}}, {381 ({1994}).

\bibitem{Holroyd2002}
H.~C. B. and C.~M. G.H.,
\newblock Psychological Review {\bf 109}, 679 (2002).

\bibitem{Shiffrin1977}
S.~R. M. and S.~Walter,
\newblock Psychological Review {\bf 84}, 127 (1977).

\bibitem{John2000}
W.~A. T.-S. John J.~McDonald and S.~A. Hillyard,
\newblock Nature {\bf 407}, 906–908 (2000).

\bibitem{Meijer1992}
P.~Meijer,
\newblock Biomedical Engineering, IEEE Transactions on {\bf 39}, 112 (1992).
}}}}}
\end{thebibliography}
\end{document}